\documentclass[twocolumn,showpacs,preprintnumbers,amsmath,amssymb]{revtex4}
\usepackage{dcolumn}
\usepackage{bm}

\newcommand{\be}{\begin{equation}}
\newcommand{\ee}{\end{equation}}
\newcommand{\bea}{\begin{eqnarray}}
\newcommand{\eea}{\end{eqnarray}}

\newcommand{\gs}{\ensuremath{g_s}} 
\newcommand{\ls}{\ensuremath{l_s}} 

\newcommand{\tr}{\mathop{\rm Tr}}
\def\expec#1{\langle #1 \rangle}

\newcommand{\cO}{{\mathcal{O}}}
\newcommand{\cN}{{\mathcal{N}}}

\newcommand{\bS}{{\mathbf{S}}}
\newcommand{\bT}{\ensuremath{\bar{T}}}

\newcommand{\aD}{\ensuremath{\overline{\mbox{D3}}}}

\newcommand{\DD}{\ensuremath{\mbox{D-}\overline{\mbox{D}}}}
\newcommand{\bN}{\ensuremath{\bar{N}}} 
\newcommand{\Et}{\ensuremath{M_{\mathrm{FT}}}}  
\newcommand{\Eg}{\ensuremath{\mathcal{E}_{\mbox{\small g}}}}  
\newcommand{\Sm}{\ensuremath{S_{\mathrm{FT}}}}
\newcommand{\Ss}{\ensuremath{S_{\mathrm{SG}}}}
\newcommand{\Ms}{\ensuremath{M_{\mathrm{SG}}}}
\newcommand{\Th}{\ensuremath{T_{\mathrm{FT}}}}

\begin{document}

\preprint{ICN-UNAM-02/03}

\title{Black Brane Entropy\\from Brane-Antibrane Systems}
\thanks{Invited talk by A.G. at the IV Mexican Workshop on Gravitation and Mathematical Physics,
based on the original work \cite{tachyon}, which should be consulted for a complete reference list.}

\author{Ulf H. Danielsson}
 \email{ulf@teorfys.uu.se}
\affiliation{%
Institutionen f\"or Teoretisk Fysik\\
Box 803, SE-751 08 Uppsala, Sweden
}%

 \author{Alberto G\"uijosa}
 \email{alberto@nuclecu.unam.mx}
\affiliation{
Departamento de F\'{\i}sica de Altas Energ\'{\i}as\\
Instituto de Ciencias Nucleares, UNAM\\
Apartado Postal 70-543, M\'exico, D.F. 04510
}%

\author{Mart\'{\i}n Kruczenski}
 \email{martink@physics.utoronto.ca}
\affiliation{
Department of Physics,
University of Toronto\\
60 St.~George~st., Toronto ON, M5S 1A7 Canada
}%

\date{April, 2002}

\begin{abstract}
In the context of string theory, it is possible to explain the microscopic origin of the
entropy of certain black holes in terms of D-brane systems. To date, most of the cases studied in
detail refer to extremal (supersymmetric) or near-extremal black holes. In this work we propose a
microscopic model for certain black branes (extended versions of black holes) which would apply to
cases arbitrarily far from extremality, including the Schwarzschild case. The model is based on a
system of D-branes and anti-D-branes, and is able to reproduce several properties of the
corresponding supergravity solution. In particular, the microscopic entropy agrees with
supergravity, except for a factor of $2^{p/p+1}$, where $p$ is the dimension of the brane.
\end{abstract}

\maketitle

\section{Strings, D-branes, and Entropy}
\label{1sec}

String theory replaces point particles with strings, one-dimensional objects whose tension
$T_F\equiv 1/2\pi\ls^2$ defines a dimensionful parameter $\ls$, known as the string length.
Besides moving as a whole, a closed string can oscillate in different ways, and upon quantization
these internal modes give rise to a perturbative spectrum consisting of an infinite tower of states
with masses
$m^2=4N/\ls^2$, $N=0,1,\ldots$ At the bottom of the tower there are massless states,
corresponding to a graviton $h_{\mu\nu}$, an antisymmetric tensor field $B_{\mu\nu}$, a scalar
$\varphi$ (the dilaton) whose vacuum expectation value determines the string coupling constant
$\gs=\exp(-\varphi)$, the so-called Ramond-Ramond (R-R) gauge fields $C_{\mu_1\ldots\mu_{p+1}}$ for various values of $p$, and the accompanying superpartners. At low energies
($E\ll \ls^{-1}$) these are the only relevant modes, and the effective field-theoretic description
is in terms of ten-dimensional Type II supergravity, with Newton's constant $G_N\sim\gs^2\ls^8$.

The non-perturbative spectrum of string theory contains extended objects of various dimensions,
collectively known as branes. Particularly important among these are the D$p$-branes \cite{polchrr},
solitonic objects extended along $p$ spatial dimensions, whose tension is inversely proportional to
$\gs$. The excitations of a D-brane are described by open strings with
endpoints constrained to lie on the brane; quantization of these strings gives rise to another
infinite tower of states. At the massless level one obtains a $(p+1)$-dimensional gauge field
$A_{\alpha}$, $9-p$ scalars $\Phi_i$ describing oscillations of the brane in the transverse
directions, and superpartners. An important point is that, in the presence of $N$ parallel
D$p$-branes, there are $N^2$ different types of open strings (each string can start and end on any
one of the branes), which means that the gauge field and scalars become $N\times N$ matrices. At low
energies, this system is described by $(p+1)$-dimensional $U(N)$ super-Yang-Mills theory (SYM) with
16 real supersymmetries and coupling constant $g_{YM}^2~\sim\gs\ls^{p-3}$. If $N$ is large, the
strength of the interactions is in effect controlled not by $g_{YM}^2$, but by the 't~Hooft coupling
$g_{YM}^2 N\propto \gs N$.

The other key property of D-branes is that they couple to closed strings: an open string attached to
the branes can close, and then wander off into the bulk of ten-dimensional spacetime. This means in
particular that D$p$-branes are sources for the supergravity fields, and so a large collection of
them should be describable, at low energies, as a macroscopic solution of supergravity, invariant
under translations along $p$ spatial directions. Such solitonic solutions are known as black
$p$-branes \cite{hs}; they generically involve a non-trivial metric, dilaton, and R-R
gauge field $C_{\mu_1\ldots\mu_{p+1}}$, all expressed in terms of a harmonic function
$H(r)=1+(R/r)^{7-p}$ in the $(9-p)$-dimensional space transverse to the brane, with $R\sim (\gs
N)^{1/(7-p)}\ls$ a characteristic length scale. The geometries in question have an asymptotically flat
region at large $r$, which connects at $r\sim R$ to a `throat' extending down to a horizon at $r=0$.

Supergravity itself is only an approximate description of the low-energy physics of string theory,
and so its black brane solutions are reliable only as long as their curvature is small compared to
the string scale. This requires $R\gg\ls$, i.e., $\gs N\gg 1$. As we have seen above, perturbative
SYM is valid only for $\gs N\ll 1$, so these two alternative descriptions of the D-brane system have
mutually exclusive regimes of validity. The supergravity solution corresponding to a stack of $N$
unexcited D-branes is \emph{extremal}: its mass $M=N\tau_p V$ (with $V$ the volume spanned by the
brane) and R-R charge $Q=N$ saturate the BPS-type inequality 
$M\ge |Q|\tau_p V$, which is
implied by the supersymmetry algebra for any object carrying $(p+1)$-form charge $Q$. BPS saturation
is equivalent to the statement that the brane solution preserves some fraction of the
supersymmetries (in this case, half). 

Black branes are extended versions of black holes--- as we have said, all such solutions possess an
event horizon. This means in particular that they have an entropy related to their horizon area by
the well-known Bekenstein-Hawking formula, $S_{BH}=A_{h}/4G_N$. Ever since this formula was written
down, it has been an outstanding problem to show how this entropy arises from an explicit counting
of the microscopic states of the black object. This can in fact be done for the extremal black
$p$-branes discussed above (which upon dimensional reduction may be regarded as black holes in $10-p$ dimensions), but the agreement is trivial: the horizon area of the solutions
vanishes, corresponding to the fact that the entropy of their microscopic counterpart, an unexcited
$N$ D$p$-brane system, is zero (there is only one such state). 

To attempt to carry out a non-trivial check, we should add some energy to the system (without adding
any charge). The black brane is then \emph{non-extremal}, and has a non-vanishing horizon area. The
simplest such setup is the \emph{near}-extremal black three-brane, which is understood
microscopically as a collection of slightly excited $N$ D3-branes. On the supergravity side, the
system has a charge $Q=N$ and a mass $M=N\tau_3 V+\delta M$, where $\tau_3=1/(2\pi)^3\gs\ls^4$ is
the D3-brane tension and $\delta M\ll N\tau_3 V$. The Bekenstein-Hawking formula assigns the black
three-brane an entropy
\begin{equation}
S_{BH}=2^{5/4}3^{-3/4}\sqrt{\pi Q}V^{1/4}\delta M^{3/4}+\ldots~,
\end{equation}
where the dots denote corrections which are higher order in $\delta M$.

On the microscopic side, the system is understood to be a stack of $N$ D3-branes plus a gas of open
strings. The massless open string modes should make the dominant contribution to the entropy, and so
we can identify the excess mass with the energy of a gas of massless particles (gluons, scalars, and
superpartners): 
$\delta M = n_b (\pi^2 / 16) N^2 V T^4$, with $T$ the temperature of the gas and $n_b$ the number of
bosonic (matrix) degrees of freedom (the numerical coefficient then takes their
fermionic partners into account).
The entropy of the system is entirely due to the gas,
\begin{equation} \label{sgas}
S_{\mbox{\small g}}=n_b\frac{\pi^2}{12}N^2 V T^3=n_b^{1/4}\frac{2}{3}\sqrt{\pi Q} V^{1/4}\delta M^{3/4}~.
\end{equation}  
The functional dependence of $S_{\mbox{\small g}}$ and $S_{BH}$ on $\delta M$ and $Q$ is identical; the
numerical coefficient agrees if $n_b=6$ \cite{gkp}. This number should be contrasted with the $\gs
N\ll 1$ result, $n_b=8$, arising from the two transverse polarizations of the gauge field
$A_{\alpha}$ and the six scalars $\Phi_i$.  Since the supergravity result applies only in the $\gs
N\gg 1$ regime, this discrepancy is not a contradiction. To date, the best
understanding  of this case comes from the AdS/CFT correspondence \cite{malda}: taking an
appropriate low-energy limit reduces the black three-brane solution to AdS$_5\times\bS^5$, and at
the same time makes $(3+1)$-dimensional $SU(N)$ SYM an exact description of the system of D-branes
and open strings. The exact equivalence thus obtained between string theory on the ten-dimensional
anti-de Sitter background and the four-dimensional gauge theory then \emph{predicts} in particular
that the SYM entropy at $g_{YM}^2 \ll 1$, $g_{YM}^2 N\gg 1$ is the same as that of a free field
theory with only 6$N^2$ bosons and the same number of fermions, instead of the naive $8N^2$. 

The first successful account of black brane entropy 
in terms of the open strings living on
D-branes was obtained by Strominger and Vafa~\cite{sv}, 
who were able to reproduce the exact entropy of certain extremal (yet finite-horizon-area) 
black holes in five dimensions through microscopic state counting in a D5-brane/D1-brane system. 
This was followed by a tremendous surge 
of activity \cite{entropy}, which led to the discovery of
the AdS/CFT correspondence \cite{malda} as a remarkable by-product,
and continues even today.  

Most of the examples where
a detailed microscopic description has been found
involve extremal and near-extremal black branes. 
These systems have positive specific heat, and their
entropy can be accounted for as in (\ref{sgas}),
using a gas of finite temperature 
living on the D-branes. 
Black branes which are far from extremality, like the ordinary
Schwarzschild black hole, are in this respect much more challenging.
Consider for example the case of a \emph{neutral}
black three-brane of mass $M$. Its entropy is given by \cite{entropy}
\begin{equation}\label{d3ent}
S_{BH}=2^{\frac{9}{4}} 5^{-\frac{5}{4}} \pi^{\frac{1}{4}}\sqrt{\kappa} 
M^{\frac{5}{4}}V^{-\frac{1}{4}}~,
\end{equation}
which we have expressed in terms of the gravitational coupling $\kappa\equiv\sqrt{G_N/8\pi}$.
If one tries to interpret this as the entropy of a gas of particles with
energy $E=M$ some well-known problems arise. First, since the 
power of $E$ is larger than one, 
a simple thermodynamical calculation shows that 
the specific heat is negative. Second,
by extensivity the exponent of $V$ is forced to be negative, 
which implies that 
the pressure $p = T\left(\partial S/\partial V\right)_M$ is negative. 
Finally, when the black brane evaporates
completely through Hawking radiation, 
all of its  excitations
disappear, meaning that the
field theory which describes it microscopically 
should have a vacuum with no
degrees of freedom (other than closed strings). 

As we will see, these same properties are in fact possessed by 
D-brane--anti-D-brane systems, which
have been much studied of late \cite{nonbps}. 
An anti-D$p$-brane ($\overline{\mbox{D$p$}}$-brane) 
is simply a D$p$-brane with reversed orientation--- it has the same
mass $M=N\tau_p V$ as the D-brane but opposite charge, $Q=-1$. 
A system with $N$ D3-branes and the same number of \aD-branes is therefore a natural candidate to
describe the neutral black three-brane \cite{tachyon}. 
We will study this system in the next section.  

Other string-theoretic attempts to give a microscopic description of Schwarzschild black holes 
have been made previously, most notably via the string/black hole 
correspondence \cite{hp}, and in the Matrix theory 
\cite{bfss} context \cite{bfks}.

\section{Brane-Antibrane Systems at Finite Temperature}
\label{2sec}

The excitations of a D3-\aD\ pair are described by open strings
whose endpoints are anchored on the branes. 
The 3-3 and $\overline{3}$-$\overline{3}$ strings
give rise to the usual massless
gauge fields and scalars $\{A_{\alpha},\Phi_{i}\}$, 
$\{\overline{A}_{\alpha},\overline{\Phi}_{i}\}$ (plus superpartners)
on the brane and the antibrane. 
The 3-$\overline{3}$ and $\overline{3}$-3 strings, on the other 
hand, yield 
a complex scalar field $\phi$ with negative mass-squared (i.e., a 
tachyon),
and additional massless fermions.
Since it originates from strings running between the brane and the 
antibrane, the tachyon is charged under the relative $U(1)$
(i.e., $A^{-}_{\mu}\equiv A_{\mu}-\overline{A}_{\mu}$),
but is neutral under the overall $U(1)$
($A^{+}_{\mu}\equiv A_{\mu}+\overline{A}_{\mu}$).

The tachyonic mass of $\phi$ expresses an instability of the system, 
the meaning of which becomes clear upon examination of the 
tachyon potential \cite{tachyonpot},
\be \label{v}
V(\phi)=2\tau_{3}\exp[-2|t(\phi)|^{2}]~,
\end{equation}
with $t$ related to $\phi$ through an error function,
\be \label{tphi}
|\phi|=\sqrt{\frac{\pi}{2}}\mbox{Erf}(|t|)~.
\end{equation}
The potential (\ref{v}) is of the `Mexican hat' type:
it has a maximum $V=2\tau_{3}$
at $\phi=0$ (the unstable `open string' vacuum)
and a minimum $V=0$
at $|\phi|=\sqrt{\pi/2}$ (the `closed string' vacuum).  
{}From the energy difference between the two vacua
we deduce that
condensation of the tachyon from the former to the latter
corresponds to the annihilation of the brane and 
the antibrane, which in particular implies the disappearance of
all open string degrees of freedom \cite{sennonbps}.

If we now consider this same system at a finite temperature, standard 
thermal field theory reasoning tells us what to expect: 
a small temperature should lead to an effective potential 
in which the location of the minimum has shifted away from 
$|\phi|=\sqrt{\pi/2}$. 
The physical reason 
for this is that moving towards $\phi=0$ can be thermodynamically 
favorable: it costs energy, but
it also reduces the mass of the tachyon and 
therefore increases the entropy of the tachyon gas.
The optimal configuration is the one that minimizes the free energy of 
the system, and this will vary with the 
temperature.

Let us now  determine whether and under what conditions
the open string vacuum, 
$t=\phi=0$,
can be a minimum of the effective potential. When this happens, 
the `tachyon' will no longer be tachyonic, and so (at sub-string-scale
temperatures) the massless fields 
will make the most important contribution to the free energy.
If one starts at $\phi=0$, then
sliding down the tachyon potential (\ref{v}) lowers the energy of 
the system, but it also gives mass to the relative 
gauge fields, and so decreases the entropy of the gas. 
We are interested in establishing which of these
effects dominates. 

Our system consists of $N$ brane-antibrane pairs, a tachyon condensate, and a 
gluon ($+$ transverse scalars $+$ superpartners) gas. Its free energy
at temperature $T=\beta^{-1}$ is given by
\begin{eqnarray} \label{f3}
F(\phi,\beta)&=&2\tau_{3}\tr e^{-2|t|^{2}}
+ \frac{4\pi}{(2\pi)^{3}}c N^{2}\beta^{-4}\\
{}&{}&\times\int_{0}^{\infty}dx\,x^{2}
\ln\left[\frac{1-e^{-\sqrt{x^{2}+\beta^{2}m^{2}}}}
{1+e^{-\sqrt{x^{2}+\beta^{2}m^{2}}}}\right]~, \nonumber
\end{eqnarray}
where $t$ is an $N\times N$ matrix, 
$m\sim|\phi|$  
is the mass given to the gluons by the Higgs effect,
and the numerical constant $c=8$ for the relevant 
8 bosonic $+$ 8 fermionic degrees of freedom. 
Starting at $\phi=0$ and 
letting a single diagonal tachyon condense
by an amount $\delta\phi$ gives mass to $N$ out of the $N^{2}$
species of particles in 
the gas, and so changes (\ref{f3}) by
\be \label{deltaf}
\delta F= -4\tau_{3}(\delta\phi)^{2}
   +\frac{1}{2}N\beta^{-2}(\delta m)^{2}~,
\end{equation}
which is positive for large enough temperature. Disregarding the
numerical constants, we thus learn that for
\be \label{Tcrit}
T\ge \frac{1}{\sqrt{\gs N}\ls}~.
\end{equation}
the open string vacuum is a minimum of the free energy (equivalently, 
a maximum of the entropy for fixed total energy),
and the brane-antibrane pairs do not annihilate. 

To arrive at (\ref{deltaf}) we have considered the mass that the 
relative gauge fields (and scalars)
acquire due to their coupling to the tachyon, 
but we can equivalently phrase the result in the opposite direction:
the second term of (\ref{deltaf}) 
represents a mass term for $\phi$ due to a
thermal expectation value for the relative gauge fields, 
$\expec{A^{-}A^{-}}_{T}\sim N T^{2}$, corresponding to a 
mass $m_{\phi}\sim \sqrt{\gs N} T$.

It is important to note
that the regime (\ref{Tcrit}) lies in the physically accessible sub-string-scale
region only for non-perturbative values of the coupling, $\gs N>1$, 
where we would expect the system to have a dual
supergravity description. 
The system under consideration becomes then
a viable candidate for the desired microscopic description
of a neutral black three-brane.

\section{Microscopic Model}
\label{3sec}

 Given the considerations of the previous sections,
the model that we consider \cite{tachyon}
 is the low-energy theory on the worldvolume of a 
 system of $N$ D3-branes and $N$ anti-D3-branes. 
 Since the number of uncondensed branes and antibranes
 can vary, the actual values are chosen to maximize 
 the total entropy of the system for fixed charge and mass.  

 The temperature 
 is assumed (and later confirmed) to satisfy
 $\frac{1}{\sqrt{g_s N}} \ll T \ll 1$ 
 in string units. 
As seen in (\ref{Tcrit}),
the first inequality is required for stability;
the second one allows us to ignore the massive open string modes. 
For the above temperature range to exist we must have
$g_s N\gg 1$, so we are necessarily
in the strong-coupling regime. 
We will take $g_s\ll 1$ to suppress closed string loops. 
Under these conditions, when a brane-antibrane pair annihilates 
the energy goes to the gas of open strings 
on the remaining branes and antibranes, rather than being emitted as closed 
strings (Hawking radiation), since the latter process is disfavored for small
$g_s$.     

Since we are trying to formulate a microscopic model 
for a supergravity solution,
the restriction to strong coupling was of course expected.
In the absence of a weakly-coupled regime, and
knowing that the theory is not supersymmetric,
the best one can do is to use plausibility arguments to determine
the entropy. 
{}From the structure of the brane--antibrane worldvolume theory 
one can show that, 
 if $\cO(N^2)$ fields 
are excited on the branes and antibranes, 
then the tachyons  
and the fermions acquire a mass of order 
$\sqrt{g_s N} T$,
which is in fact what we obtained (for the tachyon) in  Section 
\ref{2sec}. 
If such is the case, 
the fact that (for $g_s N\gg 1$) the temperature is much lower than 
their mass means that these fields are not excited,
and the theory on the branes decouples
from the theory on the antibranes.
We then have two copies of $(3+1)$-dimensional 
 $\cN=4$ $U(N)$ SYM. (The supersymmetries preserved by each copy 
are different, so the overall system is not supersymmetric.) 
 As we have seen in Section \ref{1sec}, at strong-coupling the 
AdS/CFT correspondence \cite{malda}
predicts that the entropy of each copy
is the same as that of a 
free field theory with $6N^2$ bosons and the same number of
fermions \cite{gkp}.  

 Given this set of assumptions, 
 the energy and entropy of our microscopic system
 follow as
\bea
 \Et &=& 2N\tau_{3}V + n_{b}\frac{\pi^2}{8}N^2 V T^4~,  
       \label{eq:EQ=0}\\
 \Sm &=&n_{b}\frac{\pi^2}{6} N^2 V T^3~,  \label{eq:SQ=0}
\end{eqnarray}
where $n_b=6$ is the number of bosonic degrees of freedom. 
Notice that, while the brane/antibrane tension and the energy 
of the two gases contribute with the same sign to $\Et$, and therefore to 
the time-time component of the energy-momentum tensor,
$T_{00}=\mathcal{E}_{\mbox{\scriptsize D}} + \Eg$, they contribute with opposite signs to the pressure: 
$T_{ij} = (-\mathcal{E}_{\mbox{\scriptsize D}} + \frac{1}{3} \Eg)\delta_{ij}$ (remember that 
$\mathcal{E}_{\mbox{\scriptsize D}}\equiv 2N\tau_3$ 
arises from the potential of a scalar field).

{}From (\ref{eq:EQ=0}) and (\ref{eq:SQ=0}) we obtain
\be
\Sm = n_{b}^{\frac{1}{4}}\frac{2^{\frac{5}{4}}}{3} \sqrt{\pi} \sqrt{N} 
       V^{\frac{1}{4}} \left(\Et - 2NV\tau_{3} \right)^{\frac{3}{4}}~,
\label{eq:S2Q=0}
\end{equation}
which is maximized by $N=\Et/5\tau_{3}V$.
 This is easily seen to imply that the energy contained in 
 the gases 
 is $3/2$ of the total tension of the branes and antibranes,
 a prediction which will be shown below to agree with supergravity.
In addition, we find that 
the temperature is $T\sim (g_s N)^{-\frac{1}{4}}$,
which as required satisfies $\frac{1}{\sqrt{g_s N}} \ll T \ll 1$.   
 Plugging the equilibrium value of $N$ back
 into (\ref{eq:S2Q=0}) we obtain the entropy-energy relation
\be
\Sm = n_{b}^{\frac{1}{4}}2^{\frac{5}{4}} 3^{-\frac{1}{4}} 5^{-\frac{5}{4}}   
      \pi^{\frac{1}{4}} \sqrt{\kappa}  
      V^{-\frac{1}{4}} \Et^{\frac{5}{4}}~,
\label{eq:SEftQ=0}
\end{equation}
where we have expressed the D3-brane tension in terms
of the gravitational coupling constant, $\tau_{3}=\sqrt{\pi}/\kappa$.

 To compare with supergravity, we recall that a neutral black three-brane 
 with Schwarzchild radius $r_0$ has mass and entropy \cite{entropy} 
\bea
\Ms &=& \frac{5}{2}\frac{\pi^3}{\kappa^2}r_0^4 V~,\\
\Ss &=& \frac{2\pi^4}{\kappa^2}r_0^5 V~,
\end{eqnarray}
which implies that
\be
\Ss = 2^{\frac{9}{4}} 5^{-\frac{5}{4}} \pi^{\frac{1}{4}}  
    \sqrt{\kappa} V^{-\frac{1}{4}} \Ms^{\frac{5}{4}}~.
\end{equation}
 Identifying $\Ms=\Et$, we see that the functional form of the 
 supergravity and field theory entropies agree.  
 With $n_{b}=6$,
 the numerical coefficient does not quite agree:
 the field theory entropy is a factor of $2^{3/4}$ too small, 
 $\Ss=2^{3/4}\Sm$. Equivalently, one can say that the supergravity 
 entropy behaves as if the gases 
 carried twice the available energy.
 
 Another interesting check is to consider the energy-momentum tensor.  
 {}From the asymptotic value of the 
gravitational field one finds that \cite{dmrr}  
$T_{ij} = -\frac{1}{5}T_{00} \delta_{ij}$. 
Putting $T_{00}=\mathcal{E}_{\mbox{\scriptsize D}} + \Eg$ and
$T_{ij} = (-\mathcal{E}_{\mbox{\scriptsize D}} + \frac{1}{3} \Eg)\delta_{ij}$,
as discussed above, one obtains $\Eg = \frac{3}{2} \mathcal{E}_{\mbox{\scriptsize D}}$, in 
agreement with the field theory prediction. 

 Before moving on to the charged case, let us discuss an 
 interesting issue that appears already here. 
 Since we reproduce the black brane entropy, 
it is clear that the specific heat of our system is negative. 
To understand why, let us consider how Hawking evaporation proceeds in this 
model, and check that the temperature increases when energy is radiated. 
When a closed string is emitted,
energy is taken from the open string gas, 
so a priori the temperature 
should decrease. However, we have found that, in equilibrium, 
the energy in the gas is $3/2$ the tension of the \DD\ pairs. This 
means that, when the gas 
has lost enough energy so as to match $3/2$ the tension 
of $N-1$ pairs, it will be entropically favorable
for one pair to annihilate, giving energy to the gas 
and increasing its temperature. 
This is repeated again and again,
effectively increasing the temperature on average as the mass of the 
system decreases. 
The process will continue until $g_s N\sim 1$, 
where the gas temperature becomes of order one in string 
units, and the model (\ref{eq:Eft})-(\ref{eq:Qft}), 
based only on the massless open string modes,
ceases to be valid.  At this point we would expect all of the 
available energy to go into a highly-excited
long string, so the brane-antibrane model makes
contact with the string/black hole correspondence
\cite{hp}.  
      
It is straightforward to generalize the model to the 
case of a charged black brane, where
the numbers of D3-branes and 
anti-D3-branes are no longer equal, $N\neq\bN$:
\bea\label{eq:Eft}
 \Et &=&  (N+\bN)\tau_{3} V 
   + E_{\mbox{\small g}}+ E_{\overline{\mbox{\small g}}}~,  \\    
 \Sm &=& 2^{\frac{5}{4}} 3^{-\frac{3}{4}} \pi^{\frac{1}{2}} V^{\frac{1}{4}} 
     \left(E_{\mbox{\small g}}\sqrt{N}+E_{\overline{\mbox{\small g}}}\sqrt{\bN}\right), 
      \label{eq:Sft}  \\
 Q_{\mathrm{FT}} &=& N - \bN~,  \label{eq:Qft} \\
T^{ii}_{\mathrm{FT}} &=&-(N+\bN)\tau_{3}
+\frac{1}{3V}\left(E_{\mbox{\small g}}+ E_{\overline{\mbox{\small g}}}\right)~,
\label{eq:Tijft}
\end{eqnarray}
where $E_{\mbox{\small g}}=3\pi^2N^2 V T^4/8$ 
($E_{\overline{\mbox{\small g}}}=3\pi^2\bN^2 V \bT^4/8$)
is the energy of the gas on the branes (antibranes).
Notice that, since the theories are decoupled, 
 the two gases
 could a priori have different temperatures, $T\neq\bT$. 

 Let us now consider the supergravity
expressions, following a procedure similar to \cite{hms}. 
 The energy, entropy and charge of a non-extremal three-brane are given by 
\bea
 \Ms &=& \frac{\pi^3}{\kappa^2} r_0^4 V \left(\cosh2\alpha 
        + \frac{3}{2}\right)~,\label{eq:Msgr} \\
 \Ss &=& \frac{2\pi^4}{\kappa^2} r_0^5 V\cosh\alpha~, \label{eq:Ssgr}\\ 
 Q_{\mathrm{SG}} &=& \frac{\pi^{\frac{5}{2}}}{\kappa} r_0^4 \sinh2\alpha~. \label{eq:Qsgr} 
\end{eqnarray}
 As discussed before, it is interesting to consider not only the mass 
 but also the other components of the energy-momentum 
 tensor, which turn out to be
\begin{equation}\label{eq:Tijsgr}
T^{ij}_{\mathrm{SG}}= \left[\frac{\pi^3}{\kappa^2} r_0^4 \left(-\cosh2\alpha 
        + \frac{1}{2}\right)\right] \delta^{ij}~.
\end{equation}
  Comparing (\ref{eq:Msgr}), (\ref{eq:Qsgr}) and (\ref{eq:Tijsgr}) with 
  (\ref{eq:Eft}), (\ref{eq:Qft}) and (\ref{eq:Tijft})  uniquely 
determines
\be
 N  = \frac{\pi^{\frac{5}{2}}}{2\kappa} r_0^4 e^{2\alpha},\ \ \ 
 \bar{N} = \frac{\pi^{\frac{5}{2}}}{2\kappa} r_0^4 e^{-2\alpha}~, 
\label{eq:NbN}
\end{equation}
 and identifies
the energy $E\equiv E_{\mbox{\small g}}+ E_{\overline{\mbox{\small g}}}$ of two the gases 
as
\be
E = \frac{3}{2} \frac{\pi^3}{\kappa^2} V r_0^4~,
\end{equation}
 in terms of which the entropy can be written as
\be
 \Ss=2^{\frac{5}{4}} 3^{-\frac{3}{4}} \pi^{\frac{1}{2}} V^{\frac{1}{4}} 
     E^{\frac{3}{4}} \left(\sqrt{N}+\sqrt{\bN}\right)~.
\end{equation}
 {}From (\ref{eq:Sft}), 
 we see that this is the correct entropy for a gas of particles 
 on the $N$ branes
plus another gas on the $\bN$
antibranes, both with the same energy $E$. 
However, since the total energy available for the gases is $E$, in the 
field theory model we have to assign an energy $E/2$ to each of 
them, resulting in a mismatch in the entropy 
which is exactly the same as found for the neutral case: $\Ss=2^{3/4}\Sm$. 
Under the condition
that the energies in both gases are the same,
one can 
check that the expressions (\ref{eq:NbN}) 
are the ones that maximize the 
entropy for fixed charge and mass. 
 
 The fact that the energy densities of 
 the two gases are the same implies that their temperatures $T$, $\bT$ 
 are different. They are related through
\be \label{2temps}
\frac{2}{\Th} = \frac{1}{T} + \frac{1}{\bT}~,
\end{equation}
where $\Th$ is the temperature defined as 
$\Th^{-1}=(\partial S/\partial M)_{Q}$ (which is a factor of
 $2^{3/4}$ smaller than the
Hawking temperature, $T_{H}=1/\pi r_o\cosh\alpha$). 
It would be interesting to try to detect these two distinct temperatures 
directly in the supergravity side, as was done for the D1/D5 system
in \cite{msgrey}.

As we emphasized before, there is nothing to prevent us from
 postulating that the temperatures of the 
 two gases are different, but
 it is not clear to us why supergravity seems to require that
 their energies (or equivalently, the pressures,) be the same.
 The equality of the two energies implies that,
as one approaches the extremal limit $M= Q\tau_{3}V$ 
(i.e., $\alpha\to\infty$ with $r_{0}\propto e^{-\alpha/2}$), 
the temperature of the gas on the antibranes 
grows without bounds. Since the model is based on massless 
open string modes, it
is expected to be valid only if both $T$ and $\bT$ are
substantially lower than the string scale. We expect
that as $\bT\to \ls^{-1}$, the 
energy available to the  gas on the antibranes
goes to a highly excited long string, whose contribution to the 
entropy is however negligible compared to the gas on the branes. 

As reviewed in Section \ref{1sec}, 
in the near-extremal region it is known that the black brane 
entropy can be precisely reproduced using a 
system without antibranes \cite{gkp}, so as one approaches extremality
one would intuitively expect a transition to this class of states. 
It is easy to see that the entropies of these two 
microscopic descriptions cross when $M\simeq 6 Q\tau_{3}V$, which 
indeed suggests a transition, with the 
brane-antibrane system being the preferred one further away from 
extremality.
However, it is hard to see how one could retain the agreement with 
supergravity in the entire parameter space: the brane-antibrane model 
gives the exact dependence of the black brane entropy on $M$ and $Q$, 
but gives a numerical value which is
always a factor of $2^{3/4}$ too small, whereas the model of 
\cite{gkp} is in accord with supergravity in the 
near-extremal region, but deviates significantly from it
already at the presumed transition point.

 Another important property of black branes is that they are 
 generically unstable \cite{gl}.  
  If the black three-brane, for instance, 
  lives on a very large torus, then it has lower entropy 
  than a ten-dimensional black hole, so the latter is the
  preferred configuration.   
 It can be shown \cite{tachyon} that in our microscopic model it is also convenient to 
  reduce the size of the system, but only until it is of the order of 
  the inverse temperature. The  
  radius that maximizes the entropy is found to yield an entropy-energy 
  relation 
which is precisely that of a black hole in 
  ten dimensions, as expected from the supergravity side. 

Before closing, we should note that 
entirely analogous models can be formulated \cite{tachyon} for the black 
two-brane and five-brane of eleven dimensional supergravity, where 
the microscopic description is based
upon M2-$\overline{\mbox{M2}}$ and M5-$\overline{\mbox{M5}}$ systems
in M-theory.
Using again the input from AdS/CFT, the results are in complete 
parallel with those of the D3-$\overline{\mbox{D3}}$ case.

\section{Conclusions}
\label{5sec}

We have shown that a \DD\ system is stable
if the temperature satisfies $T>1/\sqrt{g_s N}\ls$ 
(which in turn requires $g_s N > 1$).
  Building upon this observation and previous work \cite{hms}, 
  we have formulated an explicit microscopic model 
  for the black three-brane of ten-dimensional type IIB
string theory. Combined with input from the AdS/CFT
correspondence, the model yields an entropy which reproduces
 the supergravity result, 
 up to a puzzling factor of $2^{3/4}$.  
 Since the AdS/CFT correspondence uses 
 supergravity (in the near-extremal region), the agreement might appear 
 to be merely a consistency check, 
 but the fact is that we are describing a very 
 different regime (including the Schwarzschild case),
 and a different situation, since the number of branes 
 can change. As we have seen, our
field theory model
possesses several appealing features
which are in close correspondence with the properties
of black branes and black holes 
in supergravity.
Beyond the immediate task of questioning its assumptions and 
attempting to resolve the numerical discrepancy in the entropy,
we believe that there are several interesting aspects of the model which
merit further study.

\section{Acknowledgements}

AG would like to thank the organizers of the IV Mexican Workshop on 
Gravitation and Mathematical Physics
for the invitation to present this work.  
UD is a Royal Swedish Academy of Sciences
Research Fellow supported by a grant from the Knut and Alice Wallenberg
Foundation and by the
Swedish Natural Science Research Council (NFR). 
The work of AG is supported by a repatriation grant from the Mexican 
National Council of Science and Technology (CONACyT).


\begin{thebibliography}{9}        
    

\bibitem{tachyon}
U.~H.~Danielsson, A.~G\"uijosa and M.~Kruczenski,
``Brane-antibrane systems at finite temperature and the entropy of black  branes,''
JHEP {\bf 0109}, 011 (2001),
{\tt hep-th/0106201}.

\bibitem{polchrr} 
J.~Polchinski,
``Dirichlet-Branes and Ramond-Ramond Charges,''
Phys.\ Rev.\ Lett.\  {\bf 75} (1995) 4724,
{\tt hep-th/9510017}.

\bibitem{hs}
G.~T.~Horowitz and A.~Strominger,
``Black strings and P-branes,''
Nucl.\ Phys.\ B {\bf 360} (1991) 197.

\bibitem{gkp}
S.~S.~Gubser, I.~R.~Klebanov and A.~W.~Peet,
``Entropy and Temperature of Black 3-Branes,''
Phys.\ Rev.\ D {\bf 54} (1996) 3915,
{\tt hep-th/9602135}.

\bibitem{malda}
J.~Maldacena,
``The large $N$ limit of superconformal field theories and supergravity,''
Adv.\ Theor.\ Math.\ Phys.\  {\bf 2}, 231 (1998)
[Int.\ J.\ Theor.\ Phys.\  {\bf 38}, 1113 (1998)],
{\tt hep-th/9711200}.

\bibitem{sv} 
A.~Strominger and C.~Vafa,
``Microscopic Origin of the Bekenstein-Hawking Entropy,''
Phys.\ Lett.\ B {\bf 379} (1996) 99,
{\tt hep-th/9601029}.
    
\bibitem{entropy}
For a review, see, e.g., 
A.~W.~Peet,
``The Bekenstein formula and string theory (N-brane theory),''
Class.\ Quant.\ Grav.\  {\bf 15}, 3291 (1998),
{\tt hep-th/9712253}.

\bibitem{nonbps}
A review can be found in, e.g.,
M.~R.~Gaberdiel,
``Lectures on non-BPS Dirichlet branes,''
Class.\ Quant.\ Grav.\  {\bf 17}, 3483 (2000),
{\tt hep-th/0005029}.

\bibitem{hp}
G.~T.~Horowitz and J.~Polchinski,
``A correspondence principle for black holes and strings,''
Phys.\ Rev.\ D {\bf 55} (1997) 6189,
{\tt hep-th/9612146}, and refs. therein.

\bibitem{bfss}
T.~Banks, W.~Fischler, S.~H.~Shenker and L.~Susskind,
``M theory as a matrix model: A conjecture,''
Phys.\ Rev.\ D {\bf 55}, 5112 (1997),
{\tt hep-th/9610043}.

\bibitem{bfks}
T.~Banks, W.~Fischler, I.~R.~Klebanov and L.~Susskind,
``Schwarzschild black holes from matrix theory,''
Phys.\ Rev.\ Lett.\  {\bf 80} (1998) 226,
{\tt hep-th/9709091}.

\bibitem{tachyonpot}
A.~A.~Gerasimov and S.~L.~Shatashvili,
``On exact tachyon potential in open string field theory,''
JHEP{\bf 0010}, 034 (2000)
{\tt hep-th/0009103};
D.~Kutasov, M.~Mari\~no and G.~Moore,
``Some exact results on tachyon condensation in string field theory,''
JHEP{\bf 0010}, 045 (2000),
{\tt hep-th/0009148};
D.~Kutasov, M.~Mari\~no and G.~Moore,
``Remarks on tachyon condensation in superstring field theory,''
{\tt hep-th/0010108};
P.~Kraus and F.~Larsen,
``Boundary string field theory of the D D-bar system,''
Phys.\ Rev.\ D {\bf 63}, 106004 (2001),
{\tt hep-th/0012198}; 
T.~Takayanagi, S.~Terashima and T.~Uesugi,
``Brane-antibrane action from boundary string field theory,''
JHEP {\bf 0103}, 019 (2001),
{\tt hep-th/0012210}.

\bibitem{sennonbps}
``Tachyon condensation on the brane antibrane system,''
JHEP {\bf 9808}, 012 (1998),
{\tt hep-th/9805170}.

\bibitem{dmrr}
S.~R.~Das, S.~D.~Mathur, S.~Kalyana Rama and P.~Ramadevi,
``Boosts, Schwarzschild black holes and absorption cross-sections 
in M  theory,''
Nucl.\ Phys.\ B {\bf 527} (1998) 187,
{\tt hep-th/9711003}.

\bibitem{hms} 
G.~T.~Horowitz, J.~M.~Maldacena and A.~Strominger,
``Nonextremal Black Hole Microstates and U-duality,''
Phys.\ Lett.\ B {\bf 383} (1996) 151,
{\tt hep-th/9603109}.

\bibitem{msgrey}
J.~Maldacena and A.~Strominger,
``Black hole greybody factors and D-brane spectroscopy,''
Phys.\ Rev.\ D {\bf 55} (1997) 861,
{\tt hep-th/9609026}.

\bibitem{gl}
R.~Gregory and R.~Laflamme,
``Black strings and p-branes are unstable,''
Phys.\ Rev.\ Lett.\  {\bf 70} (1993) 2837
{\tt hep-th/9301052}.


\end{thebibliography}
\end{document}